\documentclass[3p,12pt]{elsarticle}
\usepackage{graphicx}
\usepackage{amssymb}
\usepackage{amsmath}
\usepackage{dsfont}
\usepackage[colorinlistoftodos]{todonotes}
\usepackage{threeparttable}
\usepackage{color,soul}
\usepackage{algorithm}
\usepackage{algorithmic}
\usepackage{amssymb}
\usepackage{makecell,multirow}
\usepackage[colorlinks=true]{hyperref}
\usepackage{geometry}
\usepackage{pdflscape}
\usepackage{lineno}

\usepackage{soul}
\soulregister{\cite}7 
\soulregister{\citep}7 
\soulregister\citet7 
\soulregister\ref7 
\soulregister\pageref7 
\soulregister\title7

\begin{document}

\begin{frontmatter}

\title{Optimal control towards sustainable wastewater treatment plants based on multi-agent reinforcement learning}

\author[1,5]{Kehua Chen$^{\#,}$}
\author[1,2,3]{Hongcheng Wang$^{\#,}$}
\author[4]{Borja Valverde-Pérez}
\author[1]{Siyuan Zhai}
\author[4]{Luca Vezzaro\corref{cor2}}
\author[1,3]{Aijie Wang\corref{cor1}}

\address[1]{Key Lab of Environmental Biotechnology, Research Center for Eco-Environmental Sciences, Chinese Academy of Sciences, 18 Shuangqing Road, Haidian District, Beijing, 100085, China}
\address[2]{School of Civil \& Environmental Engineering, Harbin Institute of Technology (Shenzhen), Shenzhen 518055, PR China}
\address[3]{State Key Laboratory of Urban Water Resource and Environment, Harbin Institute of Technology, Harbin, 150001, China}
\address[4]{DTU Environment, Technical University of Denmark, Bygningstorvet, Building 115, 2800 Kongens Lyngby, Denmark}
\address[5]{Sino-Danish Center for Education and Research, University of Chinese Academy of Sciences, Beijing, China}

\cortext[cor1]{ajwang@rcees.ac.cn}
\cortext[cor2]{luve@env.dtu.dk}
\cortext[]{\# Kehua Chen and Hongcheng Wang contributed equally to this article}

\begin{abstract}
Wastewater treatment plants (WWTPs) are designed to eliminate pollutants and alleviate environmental pollution resulting from human activities. However, the construction and operation of WWTPs consume resources, emit greenhouse gases (GHGs) and produce residual sludge, thus require further optimization. WWTPs are complex to control and optimize because of high non-linearity and variation. This study used a novel technique, multi-agent deep reinforcement learning (MADRL), to simultaneously optimize dissolved oxygen (DO) and chemical dosage in a WWTP. The reward function was specially designed from life cycle perspective to achieve 
sustainable optimization. Five scenarios were considered: baseline, three different effluent quality and cost-oriented scenarios. The result shows that optimization based on LCA has lower environmental impacts compared to baseline scenario, as cost, energy consumption and greenhouse gas emissions reduce to 0.890 CNY/m$^3$-ww, 0.530 kWh/m$^3$-ww, 2.491 kg CO$_2$-eq/m$^3$-ww respectively. The cost-oriented control strategy exhibits comparable overall performance to the LCA-driven strategy since it sacrifices environmental benefits but has lower cost as 0.873 CNY/m$^3$-ww. It is worth mentioning that the retrofitting of WWTPs based on resources should be implemented with the consideration of impact transfer. Specifically, LCA-SW scenario decreases 10 kg PO$_4$-eq in eutrophication potential compared to the baseline within 10 days, while significantly increases other indicators. The major contributors of each indicator are identified for future study and improvement. Last, the authors discussed that novel dynamic control strategies required advanced sensors or a large amount of data, so the selection of control strategies should also consider economic and ecological conditions. In a nutshell, there are still limitations of this work and future studies are required. 
\end{abstract}

\begin{keyword}
Wastewater treatment \sep reinforcement learning \sep multi-objective optimization \sep sustainability


\end{keyword}
\end{frontmatter}


\section{Introduction}\label{Intro}
\subsection{Motivation}

With an increasing population and the acceleration of urbanization and industrialization, a large amount of wastewater is being produced. Wastewater treatment plants (WWTPs) have been designed to eliminate contaminants and alleviate environmental pollution of wastewater resulting from human activities. Normally, there are four phases in treating wastewater: pre-treatment, primary treatment, secondary treatment, and tertiary treatment, and each phase has plenty of technologies to choose from. From a global scale, WWTPs have positive effects in environment protection \citep{ROSSO20081468}. However, the construction and operation of WWTPs consume resources (freshwater, energy, chemicals, etc.), emit greenhouse gases (GHGs) and produce residual sludge.
WWTPs are complex to control and optimize because of high non-linearity and variation. At present, traditional control strategies driven by Single Objective Optimization (SOO) lack systematic thinking \citep{YANG201488}. The main focus of WWTPs lies on effluent quality, energy consumption or cost, ignoring a comprehensive evaluation of the impact caused by WWTPs.
For example, set-points of aeration rate are normally determined by nitrogen and BOD removal efficiency \citep{aamand2013aeration}. Nevertheless, the content of dissolved oxygen (DO) can also affect the emission of nitrous oxide ($N_2O$) \citep{castro2015effect} and chemical dosage. With the rise of environmental consciousness, the optimization towards sustainability is imperative. 

\subsection{Related work and contributions}
Model Predictive Control (MPC) utilizes an explicit model to predict future responses of a system. In wastewater treatment, \cite{holenda2008dissolved} applied MPC technique to control DO in an activated sludge process; Nonlinear MPC (NMPC) was applied in ASM2d model to achieve optimal control \citep{grochowski2016supervised}. Although MPC is a powerful method to control dynamic systems, it requires detailed dynamic models with highly nonlinear differential and algebraic equations. Additionally, traditional MPC generates policies based on open-loop control rather than closed-loop manner, and MPC could not consider the uncertainties during control \citep{ernst2008reinforcement}. Therefore, data-driven or model-free algorithms try to handle these drawbacks.

Since the release of Alpha GO \citep{silver2016mastering}, Reinforcement Learning (RL) has received much attention to optimize different processes in WWTPs. RL is a branch of machine learning, where an agent learns from interacting with the environment \citep{kaelbling1996reinforcement}. 
\cite{hernandez2012emergent} applied value-based RL in WWTP to minimize effluent ammonia and energy consumption simultaneously. The environment was modelled based on Benchmark Simulation Model No.1 (BSM1). In order to communicate accurately, operation cost was used as the metric. 
\cite{syafiie2011model} used Q-learning \citep{watkins1992q} to control oxidation process in WWTP according to oxidation–reduction potential measurement. Temporal abstraction was applied to reduce the amount of non-relevant exploration and the calculation time. The aim of the agent was to maintain oxidation-reduction potential (ORP) at specific point. The ORP level was discrete based on measurement noise.
Furthermore, the hydraulic retention time (HRT) of anaerobic and aerobic reactors was optimized by Q-learning in activated sludge process based on ASM2d model \citep{pang2019intelligent,pang2019influent}. The state was generated according to effluent COD and TP, and the agent updated the HRT under four different step-lengths. 
However, although modified Q-learning such as Deep Q-Network \citep{mnih2015human} can handle problems with continuous spaces, such off-policy value-based algorithms do not stably interact with deep function approximation and require extra exploration strategies \citep{sutton2011reinforcement}. 
Different from value-based algorithms, policy-based algorithms naturally optimize quantity of interest stably under function approximation, and exploration is integrated with policies. 
REINFORCE algorithm was used to control bioprocesses \citep{petsagkourakis2020reinforcement}. The control policy was parameterized by recurrent neural network. The agent was firstly trained off-line in a simulated environment, after transfer learning, the algorithm was applied on the true system. 
A novel policy-based algorithm, proximal policy optimization (PPO), was applied to optimize the control of pump station in WWTPs \citep{filipe2019data}. The actions were sampled from Beta distribution. Tank level and pump consumption were integrated as the reward function, therefore, the aim of the agent was to control tank level within reasonable range and reduce energy consumption.

As a cradle-to-grave or cradle-to-cradle analysis technique, numerical Life Cycle Assessment (LCA) has been integrated into mathematical optimization as objective functions. \cite{de2016feasibility}, \cite{li2018multi} and \cite{ahmadi2015process} used LCA model to solve Multi-Objective Optimization (MOO) problems, thus achieving minimization of environmental impacts, which provided more comprehensive guidance on management and design of water facilities. Nevertheless, the research of optimal control based on LCA is still at the initial stage.

\subsection{Paper overview}
This paper focuses on real-time optimal control in an activated sludge based WWTP. 
An actor-critic algorithm with multi-agents, Multi-Agent Deep Deterministic Policy Gradient (MADDPG), is applied to achieve the control of dissolved oxygen and chemical dosage in a WWTP under continuous action and state spaces. In order to obtain sustainable control strategies, various reward functions are adopted and compared.  The structure of this paper is as follows: in Section 2, the layout of the WWTP and problem formulation 
are introduced, leading to the description of the learning process. In Section 3, different scenarios are compared and discussed, limitation and future work are introduced.

\section{Methodology}\label{S:2}
\subsection{WWTP overview} \label{S:WWTP}
\begin{figure}[htbp]
    \centering
    \includegraphics[width=17cm]{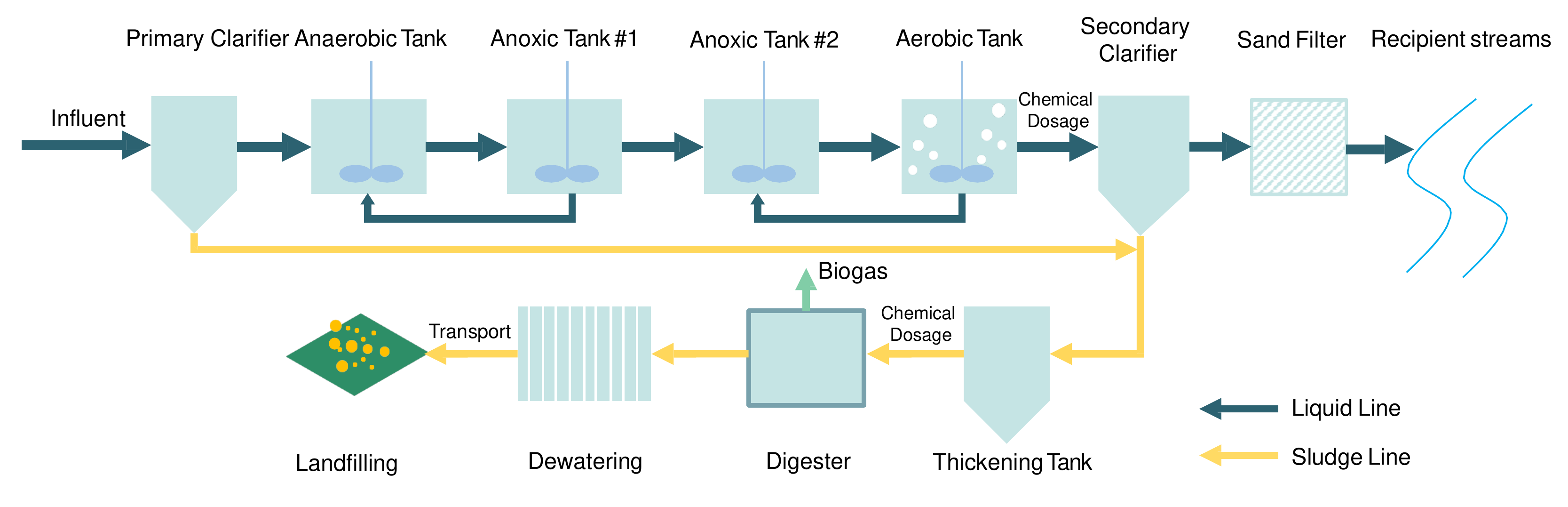}
    \caption{WWTP layout}
    \label{fig:layout}
\end{figure}

A WWTP based on activated sludge is optimized in this study as Fig. \ref{fig:layout} shows. The main process includes primary sedimentation, biological treatment, secondary sedimentation, filtration and sludge treatment. The sludge is treated by thickening, chemical dosing, digestion and dewatering. The WWTP is located in Jiangsu Province, China, with a population equivalent around 10,000.
The volumes of primary clarifier, anaerobic tank, anoxic tank 1, anoxic tank 2, aerobic tank and secondary clarifier are 300, 200, 400, 600, 800 and 600 $m^3$ respectively. The sludge recycling ratio is set as 150\%, and the internal recycling ratio (IRR) between anoxic tank 1 and anaerobic tank is 300\%, the IRR between aerobic tank 2 and anoxic tank 2 is 200\%. To further eliminate phosphorus, chemical precipitation is implemented in aerobic tank with 25\% Polyaluminium Chloride (PAC) solution. For sludge treatment, 40\% ferric chloride solution is applied for sludge pre-treatment before digestion, the dosage is 30 g/kgTSS. Water from thickening and dewatering returns back to primary clarifier. The treated sludge is transported for land-filling. 
The WWTP is simulated with MANTIS model developed by Hydromantis GPS-X. As a comprehensive model, MANTIS model integrated ASM2d, UCTADM1 \citep{sotemann2005integrated} and Musvoto precipitation model \citep{van2003modelling}.

The characteristics of influents are shown in Table \textcolor{blue}{S1}. Dynamic influent data are generated based on literature \citep{gernaey2011dynamic, flores2012benchmarking}. Since the SRT is 15 days, the first 20-day simulation is used to reach steady state, while the following 10 days are used for optimization.

Under baseline scenario, DO and chemical dosage are constant with a closed-loop control. Specifically, the DO set-point is 1.5 mg/L, and the phosphorus precipitation dosage is 0.125 kg/m$^3$-ww. The system boundary of LCA is confined within the WWTP.

\subsection{Problem Statement}\label{S:problem}
\begin{figure}[htbp]
    \centering
    \includegraphics[width=17cm]{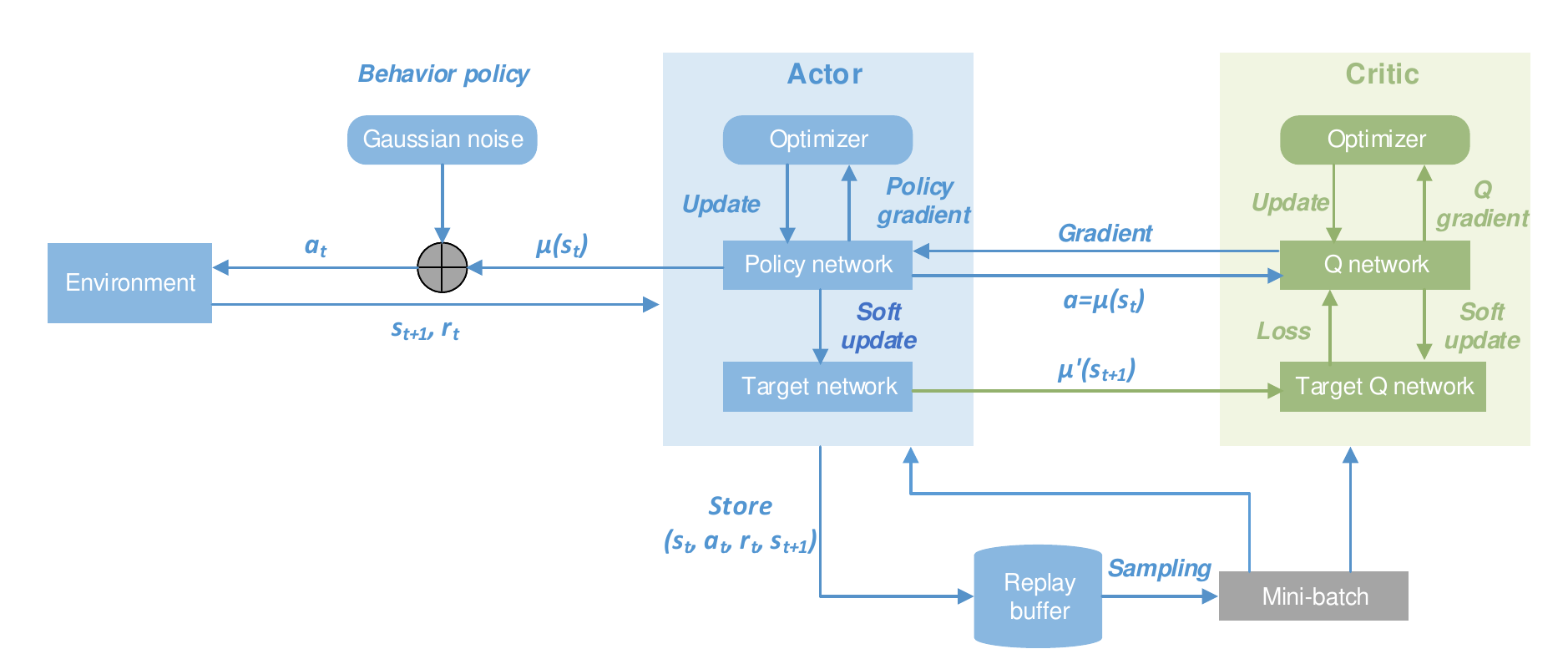}
    \caption{DDPG algorithm structure}
    \label{fig:RL}
\end{figure}

In this study, a popular actor-critic algorithm in RL, DDPG, is used \citep{lillicrap2015continuous} as shown in Fig.\ref{fig:RL}. Considering a standard RL problem, we model the environment $E$ as Markov decision process with state space $\mathcal{S}$, action space $\mathcal{A}$, an initial state distribution $p(s_1)$, transition dynamics $p(s_{t+1}|s_t, a_t)$, and reward function $r(s_t, a_t)$. 
An agent interacts with the environment by choosing different actions $a_t \in \mathds{R}^N$ at timestep $t$. After each interaction, the environment releases state $s_t$. Agent's behaviors are defined by a stochastic policy, $\pi$, which maps states to actions ($\pi: \mathcal{S}\xrightarrow{}\mathcal{P}(\mathcal{A})$). When the environment is not fully observable, state space is replaced by observation space $\mathcal{O}$.

The return is defined as the sum of discounted future reward:

\begin{gather}
    R_t = \sum_{i=t}^{T}\gamma^{(i-t)}r(s_t, a_t) 
\end{gather}
in which $\gamma \in [0, 1]$ is the discount factor. \\

The aim of RL is to find an optimal policy that maximizes expected return:

\begin{gather}
    max_{\pi(\cdot)}\mathds{E}_{r_i, s_i \sim E, a_i \sim \pi}[R_t|s_t, a_t]
\end{gather}

Many RL algorithms acquire expected return by calculating action-value functions recursively, e.g. Bellman equation:

\begin{gather}
    Q^\pi(s_t, a_t) = \mathds{E}_{r_t, s_{t+1}\sim E}[r(s_t,a_t) + \gamma \mathds{E}_{a_t\sim \pi}[Q^\pi(s_{t+1}, a_{t+1})]]
\end{gather}

In DDPG, a parameterized deterministic policy $\mu(s|\theta^\mu)$ is considered with parameter $\theta^\mu$. The critic $Q(s, a)$ is learned using the Bellman equation, while the actor is updated by the gradient of expected return from initial state with respect to actor parameters $\theta^Q$:

\begin{align}
\nabla J_{\theta^\mu} &\approx \mathds{E}_{s_t}[\nabla_{\theta^\mu}Q(s, a|\theta^Q)_{s=s_t, a=\mu(s_t|\theta^\mu)}]\\
 &= \mathds{E}_{s_t}[\nabla_a Q(s, a|\theta^Q)_{s=s_t, a=\mu(s_t)}\nabla_{\theta^\mu}(s|\theta^\mu)_{s=s_t}]
\end{align}

Similar to Deep Q-Network \citep{mnih2015human}, a replay buffer $\mathcal{R}$ is used in DDPG. Transitions are sampled from the environment by an exploration policy, and the tuple $(s_t, a_t, r_t, s_{t+1})$ are stored in the replay buffer. Gaussian noise is used for exploration. When the replay buffer is full, oldest experiences are discarded. Furthermore, soft target updates are applied in order to avoid divergence of Q update. Target networks are firstly copied from actor and critic networks, i.e. $\mu'(s_t|\theta^{\mu'})$ and $Q'(s_t, a_t|\theta^{Q'})$. Then these target networks update slowly with learned networks:

\begin{align}
\theta' \xleftarrow{} \tau\theta + (1-\tau)\theta'
 \end{align}
in which, $\tau \ll 1$.

\begin{figure}[htbp]
    \centering
    \includegraphics[width=10cm]{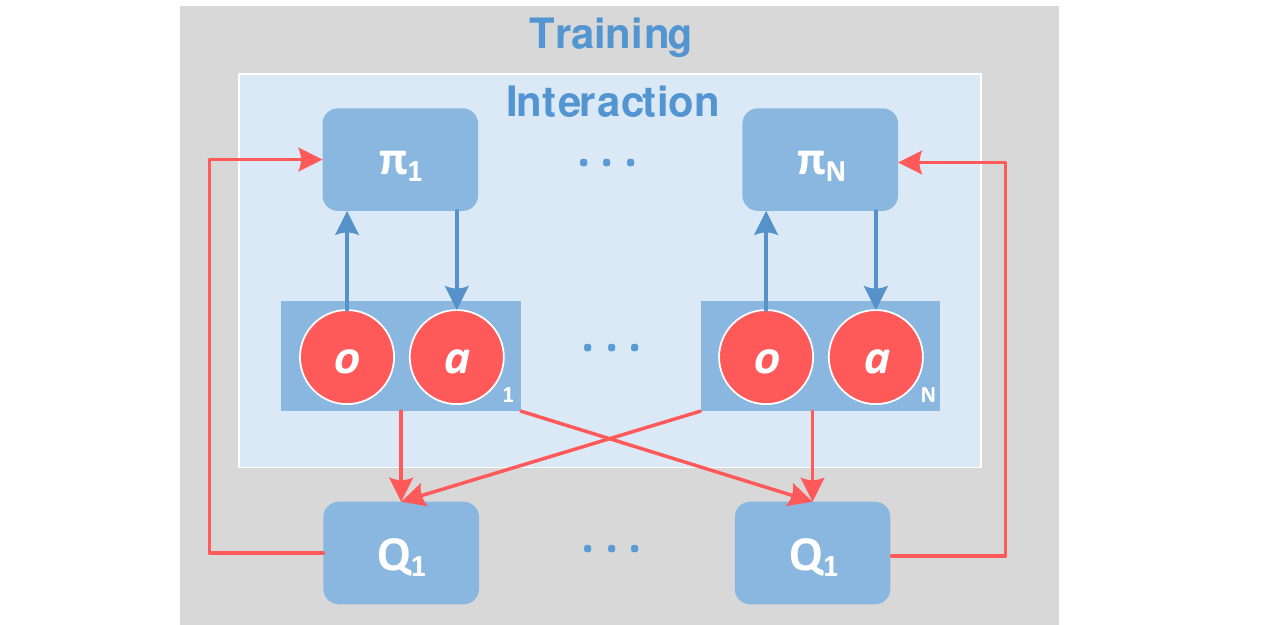}
    \caption{Overview of multi-agent structure}
     \label{fig:maddpg}
\end{figure}

In this study, two agents are deployed in terms of multi-agent paradigm, i.e. MADDPG. Compared to single-agent reinforcement learning, agents of MARL are merely able to receive local information, which is consistent to the practical control scenarios. Furthermore, previous studies proved that MARL algorithms outperformed traditional methods due to the lack of a consistent gradient signal for single-agent algorithms \citep{lowe2017multi}. Therefore, although this study only utilizes two agents for two control parameters, MARL can achieve satisfactory performance when other vital parameters are considered simultaneously, such as recycling ratio and HRT.
In detail, one agent is for DO control and the other is for dosage control. Normally, multi-agent algorithms have decentralized actor and centralized critic, which means each actor receives its own observations and outputs single actions but critic network of each agent receives complete observations (Fig. \ref{fig:maddpg}). 

In this paper, the optimization problem is abstracted to a sequential decision problem. The environment is the WWTP model as described in Section \ref{S:WWTP} coupled with interaction interface, and the agents try to set dissolved oxygen (DO) and chemical dosage in terms of deterministic policies. The observation of agents includes historical information of five timesteps: (i) influent COD, TN, TP and NH$_3$-N (in ASM state form); (ii) inflow rate; (iii) time; (iv) current DO and dosage respectively. After each interaction, a reward signal is released by the environment. The reward function is designed based on cost and LCA (see Section \ref{S:reward}). The aim of the agents is to minimize negative impacts of the studied WWTP.

\subsection{Learning process}
The MADDPG learning process mainly follows the original paper \citep{lowe2017multi} and is introduced in this section. 
Different from the original paper, Gaussian noise $\mathcal{N}$ rather than Ornstein-Uhlenbeck process is used for exploration.
Hyperparameters of MADDPG are fine tuned and listed in Table \textcolor{blue}{S2}. Before training, $10,000$ sample data are acquired by Monte Carlo sampling from uniform distribution. 
The value of DO ranges from 0 to 5 mg/L, and chemical dosage ranges from 0 to 0.5 kg/m$^3$-ww. 

\textbf{Step 0, Initialization:} Randomly initialize actor and critic weights $\theta^\mu_i$ and $\theta^Q_i$ for $i$-$th$ agent. Initialize target networks as $\theta^{\mu'}_i=\theta^\mu_i$, $\theta^{Q'}_i=\theta^Q_i$. Initialize replay buffer $\mathcal{R}$. Calculate the rewards of the sample points, obtain maximum and minimum values of each term for normalization (see section \ref{pa:weight} for details). The initial state $o$ is randomly chosen from the sample points.

\textbf{Step 1, Interaction:} The behavior policy $\beta_i$ receives the observation vector and outputs action $a_i$ to the environment. 

\begin{gather}
\label{eq:L}
    a_{t,i} = \mu(o_{t, i}|\theta^\mu) + \mathcal{N}_t
\end{gather}

The environment then interacts with GPS-X model and the agent receives a new observation $o'$. Reward $r_{t, i}$ is then calculated based on the new state, tuple $(o, a_{t, i}, r_{t, i}, o')$ is stored in the replay buffer $\mathcal{R}$. If the replay buffer is full, oldest samples will be discarded.

\textbf{Step 2, Network training:} Randomly sample N transitions (N=256 in this study) in the replay buffer as a mini-batch. For critic network, loss $L$ is calculated by mean squared loss (MSE). The critic gradient is thus $\nabla_{\theta^Q_i}L$.

\begin{gather}
\label{eq:L}
    L_i = \frac{1}{N}\sum_{j}(y^j - Q^\mu_i(o^j, a_1^j, ..., a_N^j)|\theta^Q_i))^2 \\
    y^j = r_i^j + \gamma Q_i^{\mu'}(o'^{j+1}, a_1'^j, ..., a_N'^j )|_{a_k'=\mu'_k(o_k^j)}
\end{gather}

Actor gradient is obtained from the deterministic policy gradient derived from \cite{silver2014deterministic}.

\begin{gather}
\label{eq:gradient}
    \nabla_{\theta^\mu_i}J\approx \frac{1}{N}\sum_j\nabla_{a_i} Q_i^\mu(o^j,a_1^j, ..., a_N^j)|_{a_k=\mu_k(o^j)}\nabla_{\theta_i}\mu_i(o^j)
\end{gather}

Each agent $i$ updates parameters in terms of equation \ref{eq:L} - \ref{eq:gradient}. Policy networks are updated by Adam optimizer \citep{kingma2014adam}, target networks are updated using soft updating method. After each epoch, the noise will decay 0.02\%.

\textbf{Step 3, Repetition:} If time-step reaches $T=5000$, stop; otherwise back to Step 1.


The algorithm is coded with Pytorch version 1.5 \citep{ketkar2017introduction} under Python 3.7 environment. The environment is achieved with the RL toolkit, Gym, developed by OpenAI \citep{brockman2016openai}.

\subsection{Reward function}\label{S:reward}
Life cycle cost and several LCA mid-point indicators are chosen to form the reward function respectively. Normalization is then applied to ensure a balanced evaluation in LCA reward. Since the DO and dosage mainly affect biological process and sludge production, other environmental impacts are considered as constant.

\subsubsection{Cost}
The total operational cost is divided into six components: energy cost, transportation cost, chemicals cost, sludge disposal cost, miscellaneous cost and biogas benefits. The prices are acquired from market investigation \citep{alibaba} and literature. The unit of cost in this study is CNY, and the exchange rate between CNY and USD is around 6.50:1.00 in 2021.

\begin{enumerate}[(1)]
    \item Energy cost $C_{e}$: unit price of electricity is 0.8 CNY/kWh, and consumption is derived from Section \ref{S:energy}, only direct energy consumption is considered.
    \item Chemical cost $C_{c}$: ferric chloride solution (40\%) is assumed to pre-treat sludge, with a price of 1.7 CNY/kg for FeCl$_3$(100\%); 25\% PAC solution is applied for phosphorus removal, with a price of 2.5 CNY/kg for PAC (100\%).
    \item Transportation cost $C_{t}$: unit price is 0.005 CNY/(kg$\cdot$km), the cost encompasses the transportation of FeCl$_3$(100\%), PAC (100\%) and residual sludge after treatment. Additionally, the average transport distance is 200 km.
    \item Sludge landfill cost $C_s$: residual sludge is transported for landfilling, and the treatment cost is 0.52 CNY/kg \citep{yang2015current}. 
    \item Biogas price: biogas is generated from digester for both heating and electricity production, and the subsidy of renewable energy is 0.25 CNY/kWh \citep{jiang2011review}. The specific model is demonstrated in \textcolor{blue}{Supplementary Information}.
    \item Miscellaneous cost: other costs such as labor and maintenance cost are deemed as fixed. The unit cost is 0.3 CNY/m$^3$-ww.
\end{enumerate}

The life cycle cost is the sum of six parts:

\begin{gather}
    LCCA = C_e + C_t + C_c + C_s +C_{mis} - C_{bio}
\end{gather}

where $LCCA$ is total cost in unit CNY/m$^3$-ww (m$^3$-wastewater).

\subsubsection{LCA mid-point indicators}\label{S:energy}
\paragraph{\textbf{Energy Consumption}}
Since aeration rate and dosage are the main factors are optimized and controlled in this study, energy consumed by aeration process and sludge treatment processes is included in reward function. Specifically, five components are consisted of energy consumption.
\begin{enumerate}[(1)]
    \item Dissolved oxygen is controlled by aeration mechanical power, with a fixed set-point, aeration power $p_a$ is controlled in terms of influent quality by a Proportional-Integral (PI) controller, i.e. high pollutant concentrations result in high aeration power to maintain a certain DO level since the degradation process requires more DO, and vice versa.
    
    \item Energy consumption of pumps includes residual sludge pump, thickening pump, dewatering pump. The energy consumption is calculated by equation \ref{eq:pump1} and \ref{eq:pump2}.
    
    \begin{gather}
    \label{eq:pump1}
        H = H_{static} + H_{dynamic} \\
    \label{eq:pump2}
        W = \frac{\rho gQH}{1000\eta}
    \end{gather}
     where $W$ is the pump power, kWh; $\rho$ is water density, 1000kg/m$^3$; $g$ is acceleration of gravity, 9.8 m/s$^2$; $Q$ is flow rate, m$^3$/s; $H$ is pumping head, m; $\eta$ is pump efficiency, 0.7. 
     The static water head $H_{static}$ is set as 5 m, and the frictional head loss is assumed as constant, 1 m.
    
    \item WWTPs often use chemicals to remove pollutants or pre-treat sludge. Chemicals also consume energy during production and transportation. According to \cite{longo2019enerwater}, energy consumption related to iron chloride (40\%) and PAC (25\%) is 3.4 and 1.94 kWh/kg respectively. 
    
    \item Electricity generated by biogas offsets part of the total energy consumption, hence has negative contribution.
    
    \item Energy consumed by mixing, heating or other pumping processes is assumed with fixed power and pumping head as 3 W/m$^3$, 13.58 kW and 5 m. 
    
    
\end{enumerate}

In a nutshell, total energy consumption is:

\begin{gather}
    E_{tot} = E_{aer}+E_{tran}+E_{che}+E_{other}-E_{bio}
\end{gather}

where $E_{tot}$ is total cost in unit kWh/m$^3$-ww.

\paragraph{\textbf{{Eutrophication potential}}}
Eutrophication potential measures underlying nutrient discharge of the system to recipient streams by emission factors, in unit $kg PO_4$-$eq$. Emission factors from CML database \citep{cml2002cml} are shown in Table \ref{Tab:EP}:

\begin{table}[htpb]
\setlength{\belowcaptionskip}{10pt}
\caption{$PO_4$ emission factors of various substances}
\centering
\begin{tabular}{c|c}
\hline
\textbf{Item}&\textbf{Emission factor (kg PO$_4$-eq/kg)}\\
\hline\hline
\(TP\) & 3.07\\
\(COD\) & 0.022\\
\(NH_4^+\) & 0.33\\
\(NO_3^-\) & 0.095\\
\(NO_2^-\) & 0.13\\
\hline
\end{tabular}
\label{Tab:EP}
\end{table}

Thus, eutrophication potential can be derived as:

\begin{gather}
    EP = 3.07\cdot TP_{eff} + 0.022\cdot COD_{eff} + 0.33\cdot {NH_4}^+_{eff} \\\nonumber + 0.095\cdot {NO_3^-}_{eff} + 0.13\cdot {NO_2^-}_{eff} 
\end{gather}

where $EP$ is total value in unit kgPO$_4$-eq/m$^3$-ww; TP$_{eff}$, COD$_{eff}$, NH${_4^+}_{eff}$, NO${_3^-}_{eff}$, NO${_2^-}_{eff}$ represent effluent TP, COD, NH$_4^+$, NO$_3^-$ and $NO_2^-$ respectively.

\paragraph{\textbf{{Greenhouse gas emission}}}
There are three scopes in GHG emissions: process emissions, energy emissions and material emissions. Process emissions are complicated. MANTIS model in GPS-X encompasses greenhouse gas module and simulates the emission of N$_2$O and CH$_4$ \citep{goel2012implementation}. In detail, N$_2$O and CH$_4$ emitted by anaerobic tank, anoxic tank, aerobic tank and digester are considered, emissions from clarifiers and other sludge treatment processes are ignored in the model. In addition, nitrous oxide and methane emission from effluent are estimated using emission factors from IPCC \citep{eggleston20062006}.

\begin{gather}
    CH_{4eff} = BOD_{eff} \cdot B_o \cdot MCF \\
    N_2O_{eff} = TN_{eff} \cdot EF \cdot \frac{44}{28}
\end{gather}

in which, $CH_{4eff}$ is methane emission rate, kg CH$_4$/d; $BOD_{eff}$ is BOD discharged rate, kg BOD/d; $B_o$ is maximum CH$_4$ producing capacity, 0.25 kg CH$_4$/kg BOD; $MCF$ is methane correction factor, 0.035 (fraction); $N_2O_{eff}$ is nitrous oxide emission rate, kg N$_2$O/d; $TN_{eff}$ is nitrogen in the effluent discharged to aquatic environments, kg N/d; $EF$ is emission factor for N$_2$O emissions from wastewater discharged to aquatic systems, 0.016 kg N$_2$O-N/kg N; the factor 44/28 is the conversion factor of kg N$_2$O-N into kg N$_2$O.

Indirect emissions associated to energy consumption can be calculated according to a factor based on the energy mix for China \citep{wang2016comparative}. Furthermore, the electricity derived from biogas causes a negative GHG emission in the study. Emissions are derived based on various emission factors Table \ref{Tab:GHG}.

\begin{table}[htpb]
\setlength{\belowcaptionskip}{10pt}
\caption{GHG Emission factors}
\centering
\begin{tabular}{c|c|c}
\hline
\textbf{Item}&\textbf{Emission factor}&\textbf{Reference}\\
\hline\hline
\(\text{Electricity}\) & 1.17 kg CO$_2$-eq/kWh & \citep{wang2016comparative}\\
\(\text{FeCl}_3 (100\%)\) & 0.986 kg CO$_2$-eq/kg FeCl$_3$ &\citep{parraviciniak2016greenhouse}\\
\(\text{PAC (100\%)}\) & 1.182 kg CO$_2$-eq/kgPAC &\citep{de2008greenhouse}\\
\(\text{Transportation (road)}\) & 0.000192 kg CO$_2$-eq/(kg$\cdot$ km) &\citep{zhang2016research}\\
\(\text{Nitrous oxide}\) & 298 kg CO$_2$-eq/kg N$_2$O &\citep{eggleston20062006}\\
\(\text{Methane}\) & 25 kg CO$_2$-eq/kg CH$_4$ &\citep{eggleston20062006}\\
\hline
\end{tabular}
\label{Tab:GHG}
\end{table}

Here, we assume that all chemicals and sludge are transported through road, with average distance as 200 km. The equation for global GHG estimation is:

\begin{gather}
    GHG_{tot} = GHG_{pro}+GHG_{energy}+GHG_{material}-GHG_{biogas}
\end{gather}

where $GHG_{tot}$ is total value in unit kgCO$_2$-eq/m$^3$-ww.

\paragraph{\textbf{{Weighted sum}}} \label{pa:weight}
For LCA scenario, the reward function is derived by combining energy consumption, eutrophication potential and greenhouse gas emission. Normalization is required for the comparison of different indicators, hence we need maximum and minimum values in advance. Here, maximum and minimum values are obtained by sampling, i.e. internal normalization (Eq.\ref{eq:internal}) \citep{pizzol2017normalisation}. In detail, keeping the setting of dynamic influent, actions are randomly chosen from uniform distributions at each timestep in the 10-day simulation to collect state information. Afterwards, maximum and minimum values of each item are acquired by calculating the reward of each state.

\begin{gather}
\label{eq:internal}
    Item_{norm} = \frac{Item_{tot} - Item_{min}}{Item_{max}-Item_{min}},\\
    \nonumber \forall\ Item \in I=\{E, EP, GHG\}.
\end{gather}

Therefore, the LCA reward function is as follows:

\begin{gather}
    LCA = w_{E} E_{norm} + w_{EP} EP_{norm} + w_{GHG} GHG_{norm}
\end{gather}

where $w_E$, $w_{EP}$ and $w_{GHG}$ indicate weights of energy consumption, eutrophication potential and greenhouse gas emission, respectively. 

The doctoral thesis of \citet{Wang2013} utilized weighting method based on historical data to determine weight factors of various LCA indicators qualitatively and quantitatively. His study includes nutrient recovery, greenhouse gas emission and energy consumption indicators. Nutrient recovery indicates the influence of nitrogen and phosphorus recovered from residual sludge, and has different definition from eutrophication potential. However, since these two indicators both involve the impacts rendered by nutrients, here, we consider them to have same weights. Thus, eutrophication potential, greenhouse gas emission and energy consumption indicators have weighted coefficients as 2.017, 2.754 and 2.900 respectively in China \citep{Wang2013}. After normalization, the corresponding weights are 0.26, 0.36 and 0.38.


\subsubsection{Extra constraints}
In order to obtain reasonable results, three extra constraints are added to the reward function. The first constraint is effluent quality with specific limits. If the effluent concentrations exceed the thresholds, an extra penalty +1 will be added to the total reward. In order to keep the smoothness of the control policy, the second constraint is the difference of control variables between two timesteps after normalization. Last, different action pairs may map to one scalar reward value and vice versa. Thus, the normalized values of two control variables are added to the reward, which means the policy could eliminate such ambiguity by choosing actions with low values.


At last, the reward is multiplied by -1 to keep consistency with the maximization setting. 

\subsection{Scenario introduction}
Nowadays, the stakeholders generally optimize control strategies based on cost under the constraint of discharge standards. Recently, the Chinese government proposed the Action Plan for Water Pollution Prevention and Control \citep{ministry}, which requires all WWTPs in protected areas to meet the Grade I-A standard of effluent discharge. In some specific areas, the standards are even stricter than Grade I-A \citep{zhang2016current}, i.e. the quasi standard IV for surface water. Common standards are listed in Table \textcolor{blue}{S3}, and the quasi standard IV for surface water only covers TP, TN, COD and ammonia. 
Therefore, environmental impacts of five scenarios are compared to present the results with different emphasis:

(1) Baseline scenario, static control strategy is applied, parameters are determined according to the experience, here, DO is set as 1.5 mg/L and dosage is 0.125 $kg/m^3$-$ww$.

(2) Global sustainability scenario I-A (LCA Grade I-A), optimization based on LCA reward is implemented under Grade I-A standard.

(3) Cost scenario, optimization based on LCCA reward is implemented under Grade I-A standard.

(4) Global sustainability scenario I-B (LCA Grade I-B), optimization based on LCA reward is implemented under Grade I-B standard.

(5) Global sustainability scenario quasi standard IV for surface water (LCA SW), optimization based on LCA reward is implemented under quasi surface water IV standard.


\section{Results and discussion}\label{S:4}
\subsection{Training results}
\begin{figure}[H]
    \centering
    \includegraphics[width=13cm]{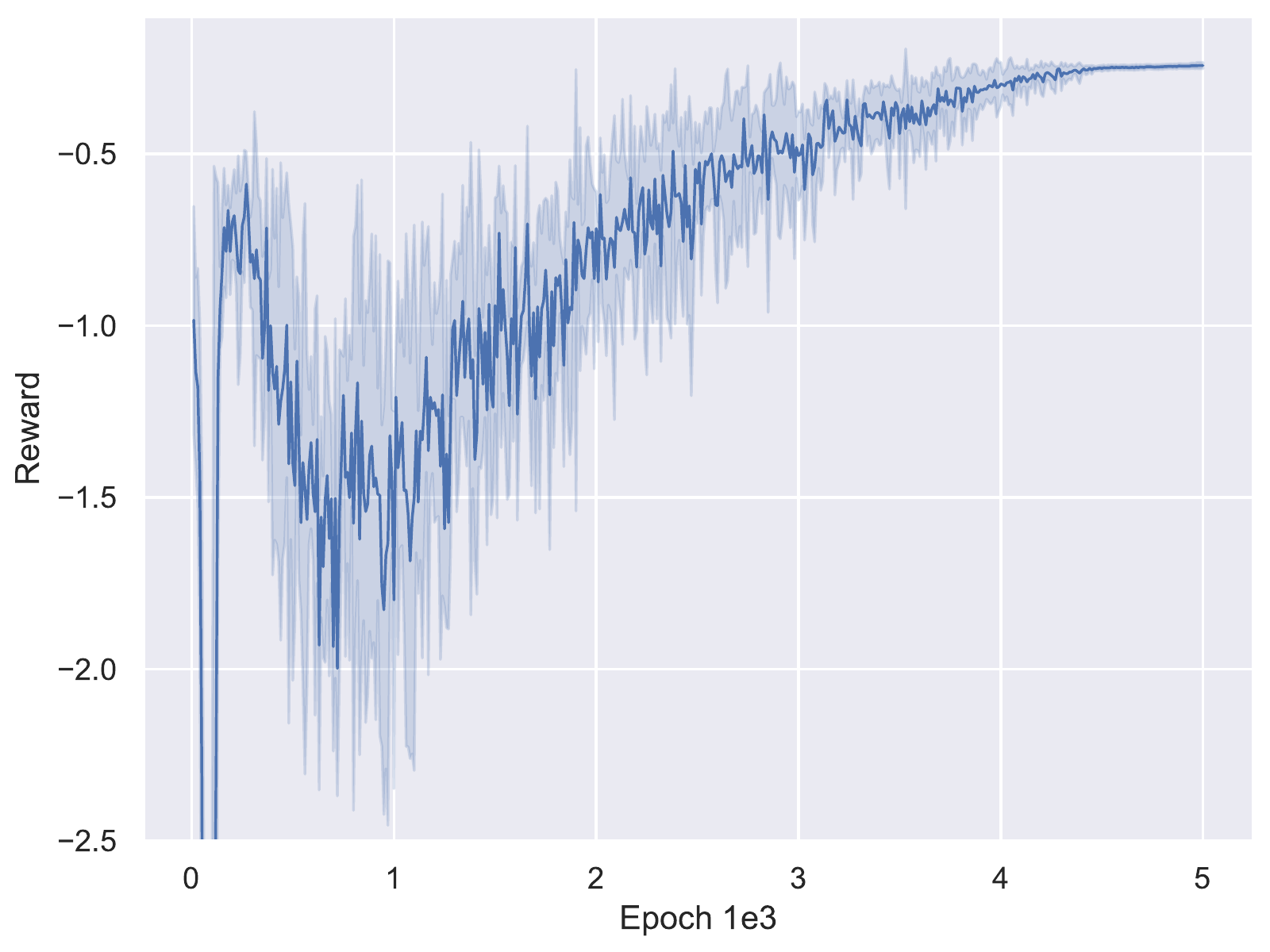}
    \caption{Reward variation under LCA scenario, the shadow is the standard deviation of 5 experiments.}
    \label{fig:reward}
\end{figure}

Fig. \ref{fig:reward} shows how the reward varies with training steps under LCA Grade I-A scenario. The shadow in the figure shows the standard deviation among five paralleled experiments.
At the beginning, the reward is low and has high variance since the initialization of weights is random and the policies have high noise to explore the action space. At the end stage, the agents learn the optimal policies and exploit the historical information sufficiently. However, since multiple value combinations lead to the similar performance, the variance still exists. In the test or deployment phase, the exploration noise changes to zero for the validation of the model efficiency. 

\begin{figure}[H]
    \centering
    \includegraphics[width=13cm]{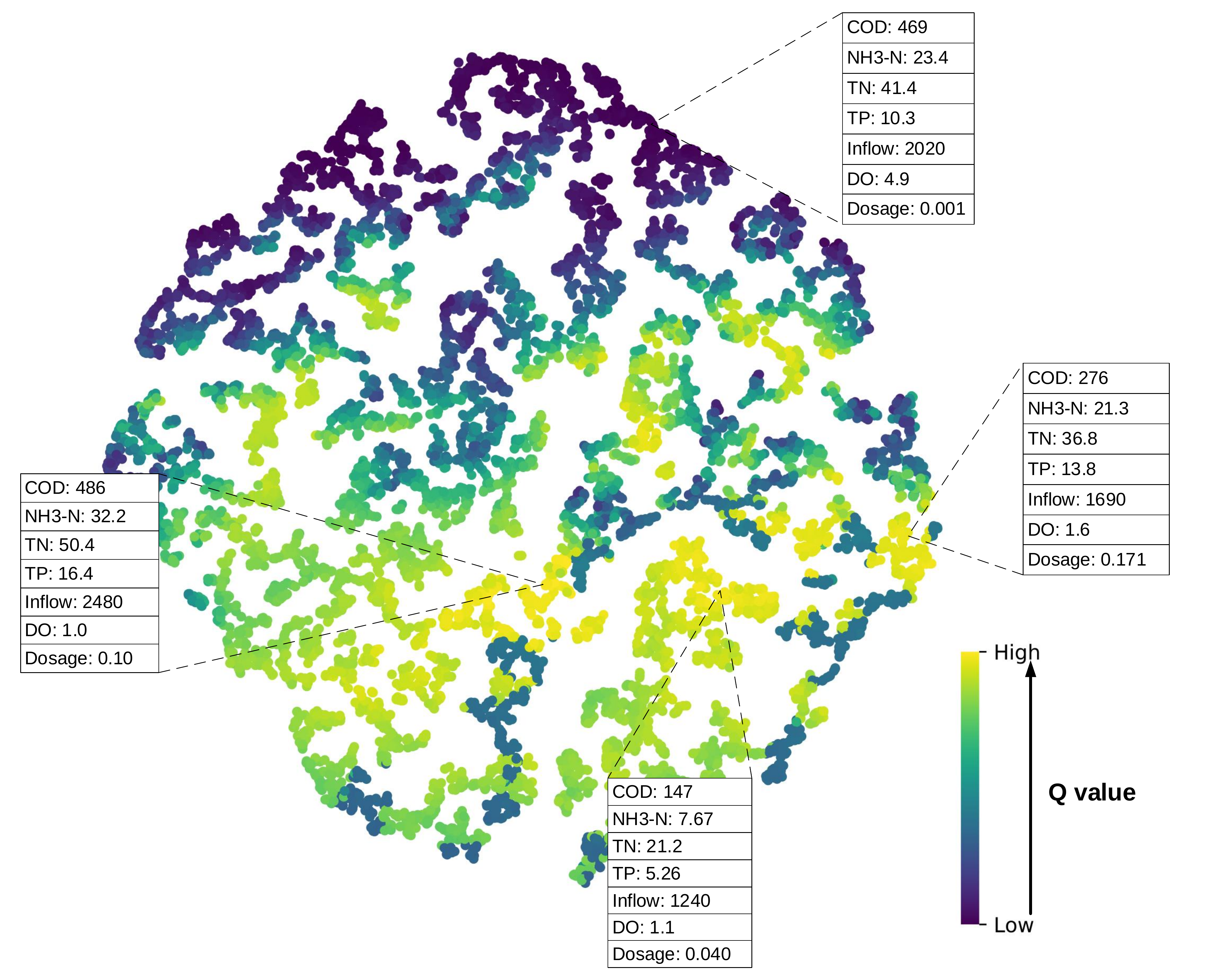}
    \caption{Two-dimensional t-SNE embedding of the input states.}
    \label{fig:tsne}
\end{figure}

The representations learned by the second last layer of critic networks are examined by using a visualization technique called 't-SNE' \citep{maaten2008visualizing} as shown in Fig. \ref{fig:tsne}. The Q values of corresponding states are generated by trained critic networks.
Generally, appropriate control parameters lead to high Q value, thus the agents tend to choose these parameters. As expected, the t-SNE algorithm tends to map the states with similar states closely. However, there are also some instances that the embedding is generated in terms of the Q values rather than the states. The reason is that neural networks are able to learn abstract features from the high-dimensional input. This conclusion is same as previous studies \citep{mnih2015human}. 

\subsection{Optimization under different scenarios}
\begin{figure}[H]
    \centering
    \makebox[\textwidth][c]{\includegraphics[width=1.15\textwidth]{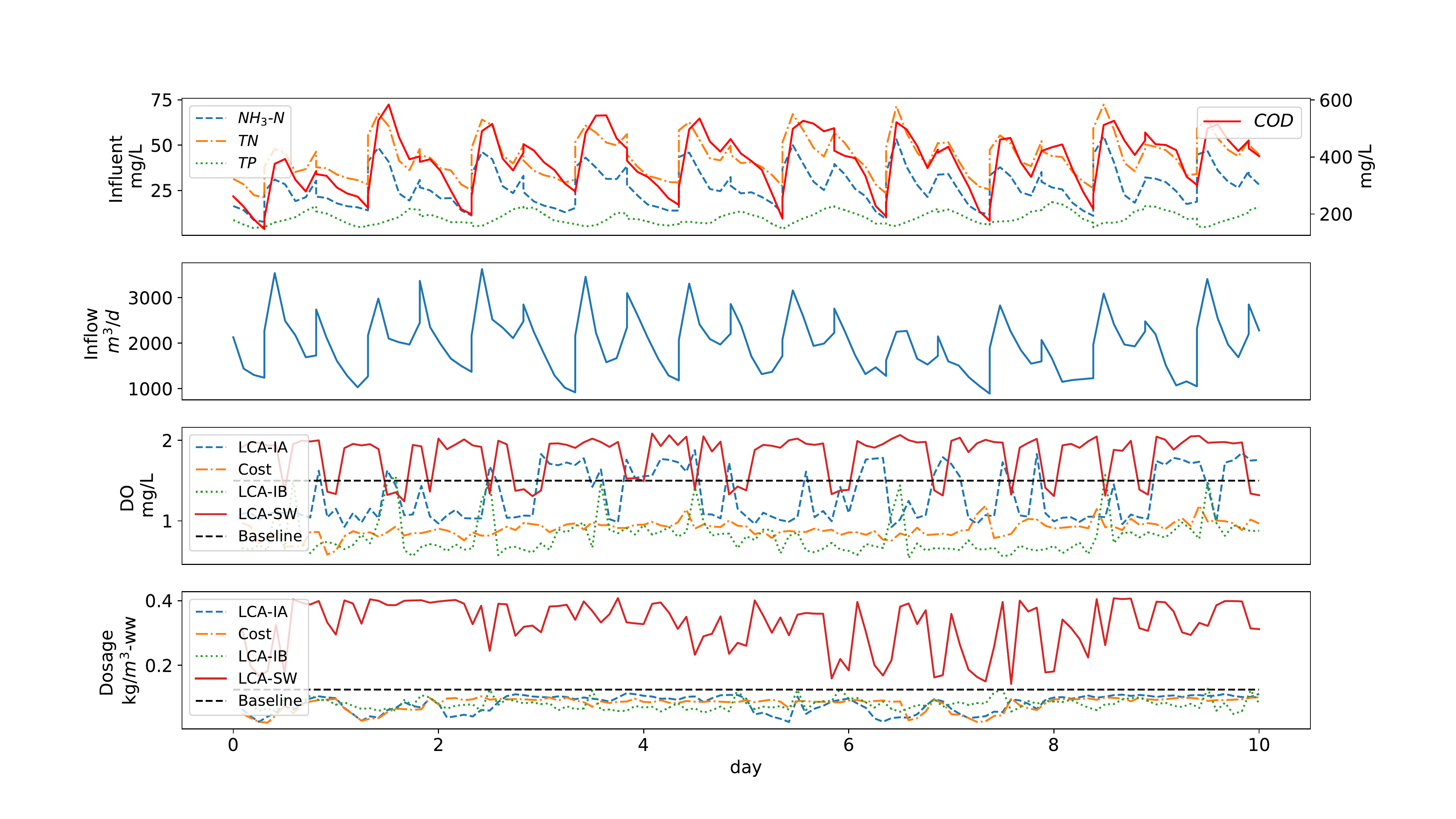}}
    \caption{Influent and optimized control parameters with time.}
    \label{fig:variable}
\end{figure}

Fig. \ref{fig:variable} demonstrates that the influent and optimized control parameters vary from day 20 to day 30. Generally, with a large inflow rate, the concentrations of pollutants decrease because of the dilution process. In addition, the concentrations and inflow rate have significant periodicity with the daily schedule.
Therefore, the optimal control variables with dynamic influent normally present a periodic variation \citep{sadeghassadi2018application}. 

Under most scenarios, the optimized DO and dosage maintain within a certain range due to the time dependence of biological system \citep{sniders2011adaptive}, if the control parameters change dramatically, the microbial community needs time to adapt or recovery from the shock. The mean DO values under LCA Grade I-A, LCA Grade I-B, cost and LCA SW scenarios are 1.3, 0.8, 0.9 and 1.8 mg/L (changed -13\%, -46.7\%, -40\% and +20\% compared to the baseline), respectively. The mean values of dosage are 0.081, 0.0796, 0.080 and 0.325 kg/m$^3$-ww (changed -35.2\%, -36.6\%, -36\% and +160\% compared to the baseline). We find that lower DO and dosage values are chosen by the agents to optimize configuration under LCA Grade I-A scenario. When the optimization only needs to meet Grade I-B standard, both DO and dosage reduce significantly. When the energy, EP and GHG are not taken into account, cost scenario decreases DO as much as possible while ensures the effluent quality. As for LCA SW, the dosage set-points show an obvious increase and fluctuation since the discharge standard of TP changes from 0.5 to 0.3 mg/L.

Fig. \ref{fig:five} presents the impacts under five scenarios along with time. The baseline has a mediocre overall performance, but steady DO and dosage set-points provide a stable environment with low variance that is beneficial to the growth of microbial community. The Chinese government requires all WWTPs constructed after 2006 meeting the Grade I-A standard, hence Grade I-A standard is considered under most scenarios \citep{dischargestandard}. Compared to the baseline, the agents optimize the system from a comprehensive perspective under LCA Grade I-A scenario. As mentioned before, both DO and dosage values decrease. As a result, the cost and energy consumption change from 0.939 to 0.890 CNY/m$^3$-ww and 0.576 to 0.530 kWh/m$^3$-ww, respectively. The DO and dosage are optimized by maintaining the effluent quality of ammonia (from 2.150 to 2.445 mg/L) and phosphorus (from 0.397 to 0.417 mg/L). In addition, optimal parameters further lower the GHG emissions compared to the baseline (from 2.527 to 2.491 kg CO$_2$-eq/m$^3$-ww) owing to the reduction of DO and dosage.

When cost-oriented optimization is implemented, the agents reduce both DO and dosage to decrease energy consumption and cost. As shown in Fig. \ref{fig:variable}, cost scenario owns lower DO rate than LCA Grade I-A scenario. 
The cost under this scenario is 0.873 CNY/m$^3$-ww. Nevertheless, the reduction in cost results in the decrease of contaminant removal efficiency, thus the eutrophication potential raised from 0.0034 to 0.0036 kg PO$_4$-eq/m$^3$-ww.

Apart from Grade I-A, Grade I-B and quasi standards IV for surface water are also taken into account. The results indicate that when the discharge standard is relaxed to Grade I-B, both energy consumption and cost reduce with the sacrifice of effluent quality (change to 0.0038 kg PO$_4$-eq/m$^3$-ww) as shown in Fig. \ref{fig:five}. The optimization under LCA SW scenario is implemented without retrofitting of the WWTP, i.e. using the same water treatment processes and technologies. Therefore, a large amount of extra DO and dosage are required to satisfy the stricter standard, and the improvement of effluent quality renders impact transfer or leakage. Compared to baseline scenario, energy consumption, cost and GHG emissions under LCA SW are 0.795 kWh/m$^3$-ww, 1.213 CNY/m$^3$-ww and 2.714 kg CO$_2$-eq/m$^3$-ww, respectively.
Besides, extra requirements are harmful to the system condition since the high variation. As a result, although the discharge pollutants reach to 0.0029 kg PO$_4$-eq/m$^3$-ww, the updating of treatment technologies is more recommended when strict standard is required.

The overall energy consumption under five cases ranges from 0.5 to 0.8 kWh/m$^3$-ww, and has consistent result with previous researches. According to the study from \citet{he2019assessment} and \cite{siatou2020energy}, the unit energy consumption decreases with the increase of WWTP scales, and small WWTPs are susceptible to technologies thus normally have high variance in values.
Moreover, the small-scaled WWTP in this study also leads to a high specific total GHG emissions as around 2.5 kg CO$_2$-eq/m$^3$-ww, while large WWTPs normally have 0.56-1.35 kg CO$_2$-eq/m$^3$-ww \citep{chen2020application}. Considering GHG emitted by biological processes, with the implementation of strict standards, the emission of N$_2$O decreases from 0.522 under Grade I-B to 0.493 kg CO$_2$-eq/m$^3$-ww under quasi surface water IV.
The reason is that high DO promoted the nitrification and reduces the nitrite concentration. According to \citet{wunderlin2012mechanisms}, nitrifier denitrification of ammonia-oxidizing bacteria (AOB) releases $\text{N}_2\text{O}$ when nitrite is added. Therefore, LCA Grade I-B scenario owns highest N$_2$O emissions.

Table \ref{Tab:cummulative} lists the cumulative environmental variations within 10 days compared to the baseline. LCA Grade I-A saves 922 kWh, 974 CNY and 722 kg CO$_2$-eq during the 10-day simulation, while cost scenario spares 1376 kWh, 1328 CNY and 1012 kg CO$_2$-eq. Therefore, the LCA Grade I-A and cost scenarios have comparable overall performance because cost scenario has higher eutrophication potential but saves more in other indicators, which means traditional cost-oriented optimization \citep{yamanaka2017total} is effective to some extent. Similarly, a relaxed standard can also lead to satisfactory comprehensive performance. As a result, the stakeholders should choose strategies according to the economic and ecological conditions. Last, LCA SW scenario affords extra 4388 kWh energy consumption, 5486 CNY cost and 3730 kg CO$_2$-eq GHG emissions within 10 days, i.e. causing great negative environmental and economic impacts.

\begin{landscape}
\pagestyle{empty}
\begin{figure}[H]
    \centering
    \makebox[\textwidth][c]{\includegraphics[width=1.8\textwidth]{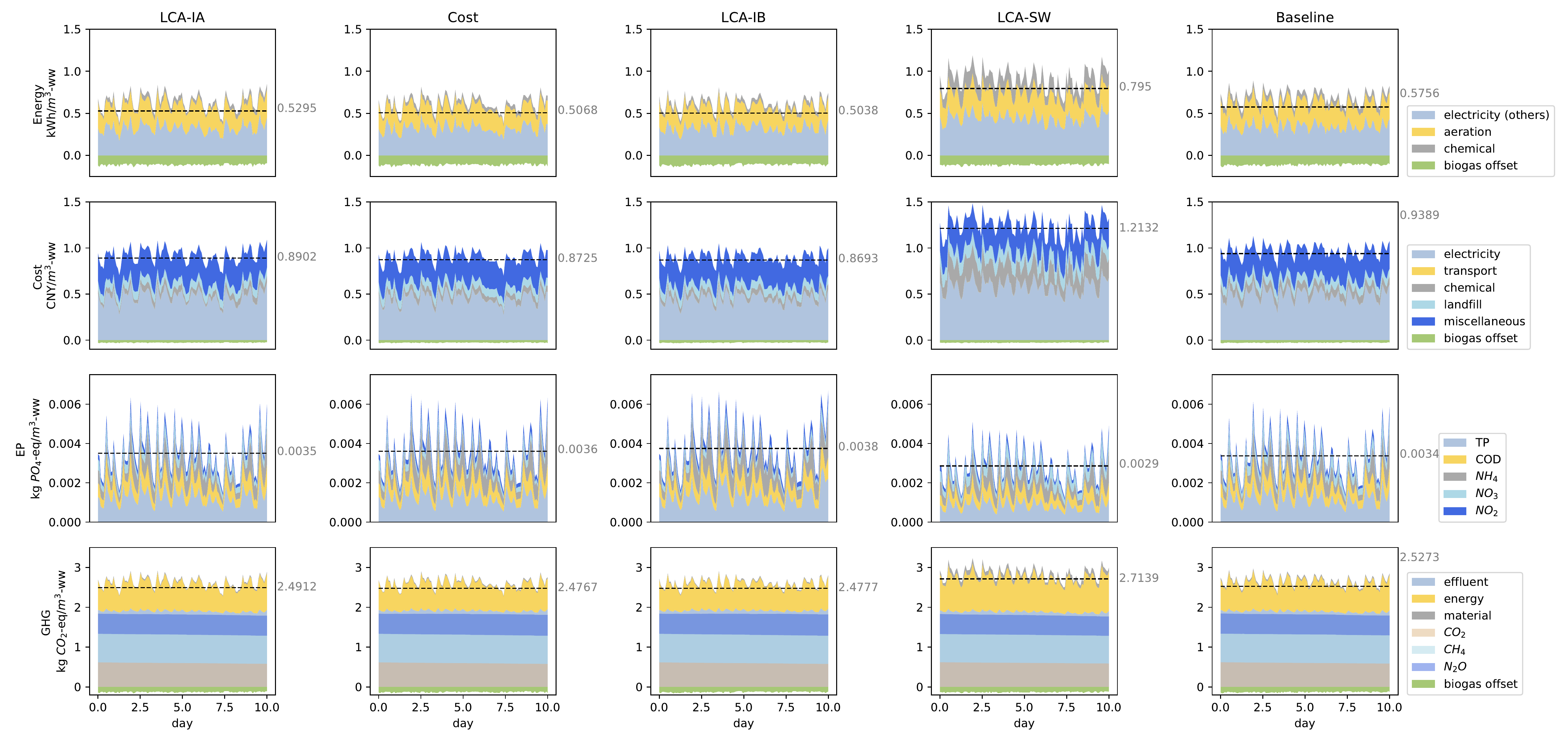}}
    \caption{LCA and cost impacts under five scenarios.}
    \label{fig:five}
\end{figure}
\end{landscape}

\subsection{Component analysis}
Among the five scenarios (Fig. \ref{fig:pie}), around 40\% of the energy consumption is caused by aeration process, which is consistent with previous studies \citep{panepinto2016evaluation}. The other electricity takes up $\sim$50\% of the total energy consumption. Generally, the chemical consumption accounts for 7-18\% of the total energy consumption. Therefore, the reduction of aeration rate and dosage is the most important and manipulable step to achieve energy saving. 

\begin{table}[htpb]
\setlength{\belowcaptionskip}{10pt}
\caption{Cumulative environmental impacts within 10 days compared to baseline}
\centering
\begin{tabular}{c|c|c|c|c}
\hline
\textbf{Scenario}&\textbf{Energy ($kWh$)}&\textbf{Cost ($CNY$)}&\textbf{EP ($kg PO_4$-$eq$)}&\textbf{GHG ($kg CO_2$-$eq$)}\\
\hline\hline
\(\text{LCA Grade I-A}\) & -922 & -974 & +2 & -722\\
\(\text{LCA Grade I-B}\) & -1436 & -1392 & +8 & -994 \\
\(\text{LCA SW}\) & +4388 & +5486 & -10 & +3730 \\
\(\text{Cost}\) & -1376 & -1328 & +4 & -1012 \\
\hline
\end{tabular}
\label{Tab:cummulative}
\end{table}

From cost perspective, electricity and miscellaneous cost account for the majority of the expenditure.
In addition, transport cost merely takes up 0.1\% of the total expenditure and is not deemed as a significant contributor.
Moreover, the cost of landfill is also non-negligible. In China, landfill is the cheapest method for sludge disposal, other methods such as incineration and building materials normally spend several times more money \citep{yang2015current}.
Hence, the sludge reduction technologies \citep{foladori2010sludge} and economical disposal approaches such as gasification \citep{gikas2017ultra} have received great attention recently \citep{gherghel2019review}.

Phosphate is the main contributor of eutrophication potential, thus, how to remove or recover phosphorus from wastewater efficiently and economically is another valuable topic \citep{wilfert2015relevance}. In all studied cases, around 60\% phosphorus is removed through biological process, and enhanced biological phosphorus removal is cost effective thus worth considering \citep{bunce2018review}. Besides, ammonia is another important factor with high eutrophication potential, hence the strict standards normally require low TP and ammonia concentrations. The analysis of microbial community shows that the concentrations of four microorganisms: heterotrophic bacteria (XBH), ammonia oxidizing bacteria (AOB), nitrifying bacteria (NOB), polyphosphate accumulating bacteria (PAO) are at similar levels under five scenarios. 
On the contrary, LCA Grade I-B scenario with lowest DO has highest PAO concentration because of the reduction of competition between PAO and XBH. As a consequence, LCA Grade I-B scenario enhances the biological phosphorus removal ($\sim$65\%). It is worth concerning that LCA SW scenario has highest biological removal efficiency as $\sim$84\% because higher DO values lead to higher microbial activity.
However, aluminum salt can inhibit biological phosphorus release and uptake processes significantly, as well as inhibit AOB dominantly \citep{liu2011inhibition}. In the simulation, the inhibition process is not considered. 

As for GHG emission, the simulation results show that GHG emitted from biological processes occupies over 70 \% of the total GHG emissions. Nonetheless, in previous studies, the emissions of secondary treatment and digestion merely covered around 50-70\% of the overall emissions \citep{flores2011including}. The reason is that the emissions of sludge disposal are included in process emissions, and additional carbon sources are not considered.
As the release of CO$_2$ is inevitable, and most CH$_4$ is collected, N$_2$O is of great importance to the global warming. N$_2$O is produced during both nitrification and denitrification processes, and the global warming potential of N$_2$O is 298 times higher than CO$_2$. \citet{kampschreur2009nitrous} concluded that low DO concentration in the nitrification stage, high nitrite concentrations, and low COD/N ratio in the denitrification stage were three main operational parameters leading to N$_2$O emissions. Hence, how to reduce N$_2$O effectively is another research topic.

From Fig. \ref{fig:pie}, we also find that the recovery of biogas is able to offset negative effects of all energy consumption, cost and GHG emissions. In this study, biogas is used for heating and electricity production, the utilization of biogas offsets around 15\% of the total energy consumption.
Nonetheless, it should be noted that the use of biogas is determined by its composition, downstream technology and policies \citep{awe2017review}. As a promising resource, high-efficient production and utilization of biogas deserve much attention.

\begin{landscape}
\pagestyle{empty}
\begin{figure}[htbp]
    \centering
    \makebox[\textwidth][c]{\includegraphics[width=1.8\textwidth]{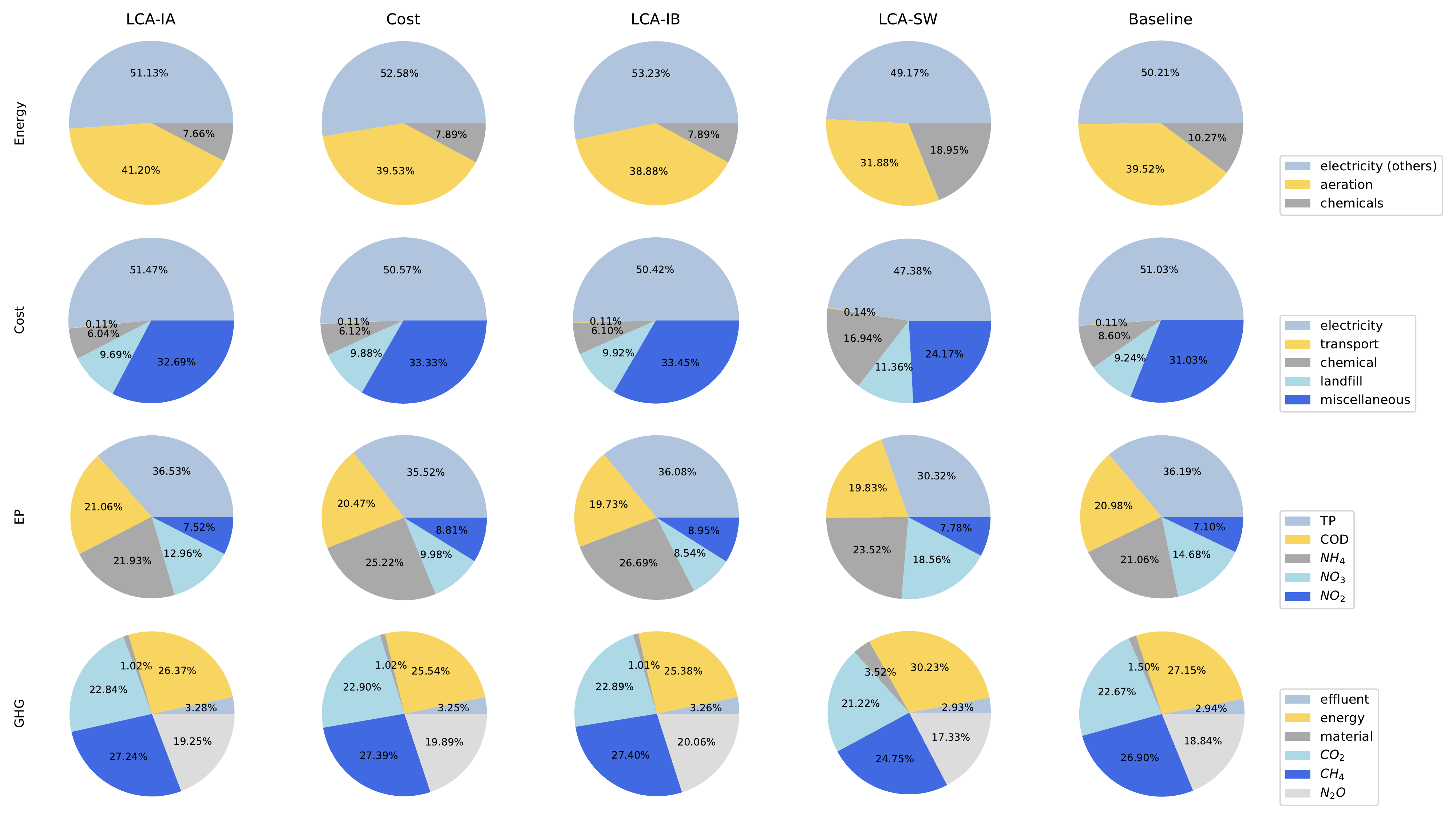}}
    \caption{Component proportion of five scenarios.}
    \label{fig:pie}
\end{figure}
\end{landscape}

\subsection{The selection of control strategies}
As discussed above, different optimization goals lead to different environmental impacts. Cost-oriented and LCA-oriented control strategies have comparable performance in this study, however, LCA-based strategies have high extendability when other environmental indicators are considered. 

Besides, if the recipient water body is sensitive to pollutants, the implementation of strict standards is imperative. Nevertheless, similar to the previous study from \citet{sweetapple2014multi}, LCA SW strategy is a case that effluent quality is considered as the dominant factor of optimization, but causes severe impact transfer. Conversely, technology-driven upgrading consumes materials and energy during construction, but may cut down side effects during operation. Here, scheme comparisons are recommended for reasonable retrofitting. 
On the other hand, if the environmental carrying capacity \citep{liu2011measurement} in the target area is high, using relaxed discharge standards (such as Grade I-B) is also an alternative.

As the most common strategy, default baseline strategies maintain control parameters at constant levels by a simple closed-loop configuration consisting of PI controllers \citep{nopens2010benchmark}, these strategies have acceptable performance, and provide stable environment for the microbial community. As for dynamic control strategies, they require real-time influent and effluent quality. There are two methods to acquire real-time data. The first method is to deploy advanced sensors \citep{campisano2013potential} requiring high expenditure for purchasing, maintenance and operating. The second one is to predict data using machine learning algorithms \citep{zhou2019random, wang2019deep}, such algorithms are trivial to achieve, but need numerous data and suffer from high uncertainty. 
In a summary, dynamic control strategies are promising but still challenging at present.

\subsection{Limitation and future work}
Multi-objective reinforcement learning is still under initiative stage \citep{liu2014multiobjective}.
In this work, a scalar reward function rather than the Pareto strategy was used. Scalar scheme is easy to understand and optimize, but cannot obtain Pareto front, which causes difficulties to strategy generation. In addition, same values of a scalar reward may indicate multiple action pairs which also brings confusion. Thus, other algorithms using Pareto framework \citep{van2014multi} can be applied in future work. Besides, RL algorithm is sensitive to reward function, hence the reward engineering can impact performance significantly \citep{dewey2014reinforcement}. This study is a trial of LCA based reward engineering, and heuristic method such as the extra constraint is imperative to avoid unrealistic results. In addition, the weights and factors impact optimization results dramatically. The determination of these parameters is according to previous literature in this study. In fact, the weighting method always involves the trade-off among political, social and ethical values, and there is no available best method. Thus, in-depth investigation is imperative for practical scenarios \citep{wang2012assessment}.
To sum up, performance of non-weighted approaches should be further investigated in future work, and an efficient reward engineering method for WWTP evaluation is also needed to be explored.

Considering that training models require plenty of time, the size of replay buffer is selected as 1000 during learning stage. Besides, DRL introduces neural networks as feature representatives, therefore, there are plenty of hyperparameters that create difficulties when acquiring good performance.

Most algorithms of RL are model-free. As data-driven algorithms, they normally require a large number of samples from the real world or models. Hence the accuracy of model determines the performance. Additionally, the dynamic GHG model is still under research stage \citep{corominas2012comparison}, and some bias is introduced by models. For example, GHG emissions from clarifiers were ignored in our GHG model; the inhibition of microbial community was not considered as mentioned above. In addition, current ASM models still have many limitations. For instance, the state variables may be not consistent to the real influent, and there are various hypothesis incorporated in models \citep{hauduc2013critical}. 
Furthermore, complicated first principle models require high computational power, which also increases the training time \citep{corominas2012comparison} and cause difficulties to online control. To reduce computational time, the surrogate model is a good alternative. 

Last, the system boundary of LCA was limited to the WWTP and merely three mid-point indicators were chosen because these indicators represented the major concerns in WWTPs and the data of these indicators was easy to acquire or calculate from models. A complete system boundary and ample indicators lead to detailed results, but also require lots of data and introduce uncertainty.

In the future, MARL can be extended to control more parameters with a more comprehensive and adaptive control policy as aforementioned in Section \ref{S:problem}. Nonetheless, how to realize effective communication and fast training process under large spaces is a tricky problem. Additionally, on-field operational data should be used to validate the novel algorithm. When the algorithm is deployed in field, the data size is usually limited, so early instruction and transfer learning can be considered to accelerate convergence. Specifically, the algorithm can be trained under simulated environment to achieve early instruction. Afterwards, field data is fed into the algorithm with the freeze of superficial layers.

With the thriving of Internet of Things, various process data can be acquired. To achieve large-scale assessments, the establishment of data centers of WWTPs is necessary in the future. This data can not only favor the process control, but also provide basis for process diagnosis and optimization.

\section{Conclusion}\label{Conc}
This study applied multi-agent deep reinforcement learning to optimize DO and dosage simultaneously. Reward functions are designed from life cycle perspective for sustainability optimization. 
The results show that: \\
(1) The novel MADDPG algorithm learns abstracts features from high-dimensional states, and influents with proper control variables have high Q values; \\
(2) Optimization based on LCA has lower environmental impacts compared to baseline scenario. The cost-oriented control strategy owns comparable performance to the LCA-driven strategy, but with less extendability; \\
(3) Environmental impacts under different discharge standards are analysed. The stakeholders should formulate standards in terms of environmental carrying capability. It is worth mentioning that the retrofitting of WWTPs based on resources should be implemented with the consideration of impact transfer; \\
(4) The major contributors of each indicators are identified for future study and improvement. Specifically, the reduction of aeration and chemical dosage lower both energy and cost significantly, while sludge disposal is another vital point for cost reduction. Enhanced phosphorus and ammonia removal further decreases the eutrophication potential. Besides, how to reduce nitrous oxide economically and effectively is another valuable topic; \\
(5) Last, we discuss that novel dynamic control strategies require advanced sensors or a large amount of data, so the selection of control strategies should also consider economic and ecological conditions. 


\section*{Acknowledgements}\label{Ackn}
We gratefully acknowledge the financial support by the Natural Science foundation of China(No.52000054), the NSFC-EU Environmental Biotechnology joint program (No.31861133001) and Shenzhen Science and Technology Program (Grant No. KQTD20190929172630447). 

\bibliographystyle{model1-num-names}
\biboptions{authoryear}
\bibliography{BibItems.bib}

\end{document}